\documentclass[fleqn,usenatbib]{mnras}

\pdfoutput=1
\usepackage{newtxtext,newtxmath}

\usepackage[T1]{fontenc}
\usepackage{ae,aecompl}

\usepackage{graphicx}	
\usepackage{amsmath}	
\usepackage{amssymb}	

\title[Masses of the Kepler-419 Planets from Transit Timing Variations Analysis]{Masses of the Kepler-419 Planets from Transit Timing Variations Analysis}

\author[Saad-Olivera et al.]{
X.~Saad-Olivera,$^{1}$\thanks{E-mail: ximena@on.br}
A.~Costa de Souza,$^{1}$
F.~Roig$^{1}$
and D.~Nesvorn\'y$^{2,1}$
\\
$^{1}$Observat\'orio Nacional, Rua Gal. Jose Cristino 77, Rio de Janeiro, RJ 20921-400, Brazil\\
$^{2}$Department of Space Studies, Southwest Research Institute, 1050 Walnut Street, Suite 300, Boulder, CO 80302, USA
}

\date{Accepted 2018 October 30. Received 2018 October 5; in original form 2018 September 14}

\pubyear{2018}

\begin{document}
\label{firstpage}
\pagerange{\pageref{firstpage}--\pageref{lastpage}}
\maketitle

\begin{abstract}
We perform dynamical fits to the Transit Timing Variations (TTVs) of Kepler-419b. The TTVs from 17 Kepler quarters are obtained from Holczer et al. (2016, ApJS 225, 9). The dynamical fits are performed using the \texttt{MultiNest} Bayesian inference tool, coupled to an efficient symplectic N-body integrator. We find that the existing TTV data alone are able to uniquely constrain the planetary masses of Kepler-419b and c. Our estimates are in a good agreement with previous mass determinations that combined different techniques and observations, such as TTVs, radial velocity measurements, and the photoeccentric effect. As expected, however, our mass estimates have larger uncertainty. We study the global stability of the system within the parameters uncertainties to discard possible unstable solutions. We conclude that our method applied to the Holczer et al. data can provide reliable determinations of the planetary parameters. In the fore-coming work, we will use Holczer et al. 2016 data to determine planetary parameters for a large set of Kepler systems.
\end{abstract}



\section{Introduction}
One of the most important goals in exoplanetary science is to characterize the masses and radii of the planets. 
In general, these parameters are difficult to obtain at once using a single observational method. For example, transit light curves allow us to directly determine the planetary radii, but not the masses. On the other hand, radial velocity (RV) measurements allow us to constrain the planetary masses, but not the radii. Therefore, it is often necessary to combine different observational methods to get a more complete characterization of planet properties. However, while transit light curves are available for a large set of stars observed by the Kepler mission, RV measurements are only available for a small sample of Kepler stars. This highlights the importance of the Transit Timing Variation (TTV) method. 

TTVs are deviations of the individual transit times of a planet from a linear ephemeris. They are often caused by planetary companions that gravitationally interact with the transiting planet  \citep[e.g.][]{Agol2005, Holman2005}. 
The TTV method explores the dynamics of the planetary system letting us constrain the masses and orbits of the transiting planets \citep[e.g.][]{Holman2010}. It may also lead to detect and characterize non transiting planets in a system where at least one planet is transiting \citep[e.g.][]{Ballard2011, Nesvorny2012,Nesvorny2013}. This is also the case of the Kepler-419 system.

\citet{Borucki2011b} determined that Kepler-419 is a K star with a mass of $\simeq1.23\, M_\odot$, and identified transits of a planet candidate that received the provisional 
designation KOI-1474.01. Later analysis by \citet{Dawson2012} showed a low false-positive rate expectations and led to estimate a planetary radius of $10.8\,R_\oplus$, and an orbital period of $P=69.7$~days. The planetary eccentricity was also constrained to be very large from the photoeccentric effect\footnote{The photoeccentric effect derives the orbital eccentricity of individual transiting planets based on the shape of the light curve and the transit duration. It takes advantage of the difference between the stellar density derived from spectroscopic observations and the stellar density inferred assuming a light curve produced by a planet in a circular orbit \citep{DawsonJohnson}.}. \citet{Dawson2012} also pointed out that the light curves of KOI-1474.01 display large TTVs, probably due to the presence of another non transiting planet, but they did not have enough TTVs observations to properly characterize it. 

Later on, \citet{Dawson2014} extended the study of Kepler-419 by including new RV measurements, which confirmed the planetary nature of KOI-1474.01, then renamed as Kepler-419b. With these new observations, the authors improved the estimates of the stellar parameters, finding a stellar mass of $1.22^{+0.12}_{-0.08}\,M_\odot$, and a stellar radius of 
$1.4^{+0.37}_{-0.21}\,R_\odot$. In order to obtain the planetary parameters, \citet{Dawson2014} first analyzed the TTVs and the RV data separately, and concluded that RVs can be used to constrain the mass of the transiting 
planet, Kepler-419b, while TTVs can be used to determine the mass of the non-transiting planet, Kepler-419c. Then, combining the available data, they fully characterized the system, confirming the very eccentric nature of Kepler-419b's orbit ($e_b\simeq0.81$), its mass ($M_b\simeq2.5\,M_{\mathrm{Jup}}$), and the mass of Kepler-419c ($M_c\simeq 7.3\,M_{\mathrm{Jup}}$).

Recently, \citet{Almenara2018} extended the study of \citet{Dawson2012, Dawson2014} by considering all available Kepler observations, as well as new RV measurements. They applied a photo-dynamical model to fit the transit light curves simultaneously with the RV curves, and obtained a full set of parameters for the two planets and the star. In particular, they estimated larger values of the stellar mass and radius ($1.39\pm 0.48\,M_\odot$ and $1.80\pm 0.22\,R_\odot$), but similar values of the planetary masses ($M_b = 2.71\pm 0.66\,M_\mathrm{Jup}$, $M_c = 7.4\pm 2.6\,M_{\mathrm{Jup}}$). Almenara et al. (2018) also showed that TTVs alone could better constrain the mass of Kepler-419b, provided that additional transits were measured.

In this work, we use Kepler-419 mid-transit times reported by \citet{Holczer2016}. We apply a Bayesian inference method to fit this data to a two-planet dynamical model. Our main goal is to use the Kepler-419 system as a test case to show that very good estimates of the planetary masses can be obtained by combining the Holczer's observed mid-transit times data set with the right statistical tool. This is of major relevance since \citet{Holczer_Catalog} catalog provides TTVs measurements for hundreds of Kepler systems for which RV observations are still not expected to be available in the near future. We also address the dynamics of the Kepler-419 system by analyzing the orbital behavior of the best fit solutions, and discuss how this poses constraints on the estimated parameters.

This paper is organized as follows: in Sect.~\ref{methods} we describe our methodology. The results are presented in Sect.~\ref{results}. Sect.~\ref{dynamic} is devoted to the dynamical analysis. Finally, in Sect.~\ref{discu}, we discuss our results in the light of those of \citet{Almenara2018}, and present our conclusions.

\section{Methodology}
\label{methods}
Our method is based on fitting the mid-transit times observed to a two-planet dynamical model with 12 
free parameters. The dynamics of the system is simulated with the \texttt{Swift} N-body code \citep{Levison1994}, 
which has been modified to determine the mid-transit times \citep{Nesvorny2013, Deck2014}. The parameter estimation 
is done with the Bayesian inference tool \texttt{MultiNest} \citep{Feroz2009, Feroz2013}, adopting a Gaussian likelihood function. The code also provides the so-called Bayesian evidence of the fit, $Z$, which allows us to perform model/solution selection. A detailed description of the method is presented in \citet{Saad-Olivera2017}. It is worth mentioning that it is better to fit mid-transit times than TTVs, because the former are directly measured from the lightcurve (or calculated from the dynamical model), while the latter have to be inferred through an intermediate fit to a linear ephemeris, which may constitute an additional source of error.

The 12 free parameters of the fit are the: planet-to-star mass ratios (${M_b}/{M_{\star}}$, ${M_c}/{M_{\star}}$), the orbital periods ($P_b$, $P_c$), the eccentricities ($e_b$, $e_c$), the longitudes of periastron ($\varpi_b$, $\varpi_c$), the mean longitude ($\lambda_c$) and impact parameter ($b_c$) of the non-transiting planet, the difference in nodal 
longitudes ($\Omega_c-\Omega_b$), and time lapse ($\delta t$) between the reference epoch ($\tau=2\,454\,959.3$~BJD) and the closest transit of Kepler-419b. This latter parameter is related to the mean longitude of Kepler-419b, 
$\lambda_b$, at the reference epoch. 

The impact parameter of the transiting planet ($b_b$) is initially set to $b=0.26$ determined from the transit light curve fit \citep{Dawson2012}. We use a transit reference system where the $x$-axis is oriented towards the observer and the nodal longitude of the transiting planet is fixed to $\Omega_b=270^\circ$ \citep[e.g.][]{Nesvorny2012}. The radius and mass of the star are also fixed, assuming two possible values: the \citet{Dawson2012} values ($R_\star=1.4\,R_\odot$, $M_\star=1.22\,M_\odot$), and the \cite{Almenara2018} values ($R_\star=1.8\,R_\odot$, $M_\star=1.39\,M_\odot$). For all the 12 model parameters, we consider uniform priors within the ranges listed in Table~\ref{tab:priors419}. These priors have been informed from the previously known information of the system \citep{Dawson2012, Dawson2014, Almenara2018}.

Orbital configurations with transits of Kepler-419c are excluded (i.e $b>1$), since no such transits have been identified in the Kepler data.

\begin{table}
	\caption{Prior distributions of 12 parameters of our model. $[x,y]$ is a uniform distribution between $x$ and $y$. Indices $b$ and $c$ refer to each of the two planets, index $p$ refers to any of the planets.}
	\label{tab:priors419}
    \centering
             \begin{tabular}{ll}
             \hline         
             \hline         
             Parameter      & Prior values  \\ [0.5ex] 
             \hline         
             \vspace{0.1cm}
             $M_b/M_\star$ & $[0.001,\,0.004]$  \\
             \vspace{0.1cm}
             $M_c/M_\star$ & $[0.003,\,0.009]$  \\
             \vspace{0.1cm}
             $P_b$ (d) & $[69.5,\,70.0]$  \\
             \vspace{0.1cm}
             $P_c$ (d) & $[600,\,800]$  \\
             \vspace{0.1cm}
             $e_b$ & $[0.8,\,0.95]$ \\                
             \vspace{0.1cm}
             $e_c$ & $[0,\,0.3]$ \\                
             \vspace{0.1cm}
             $b_c$ & $[1,\,20]$  \\
             \vspace{0.1cm}                   
             ${\varpi}_p$~($^\circ$) & $[0,\,360]$ \\
             \vspace{0.1cm}    
             $\lambda_c$~($^\circ$) & $[0,\,360]$ \\
             \vspace{0.1cm} 
             $\Omega_c-\Omega_b$~($^\circ$) & $[0,\,360]$   \\
             \vspace{0.1cm} 
             $\delta t$~(d) & $[0,\,0.06]$\\ 
             \hline
             \end{tabular}                
\end{table}

\section{Results}
\label{results}

\citet{Holczer_Catalog} provide information about 20 transits for Kepler-419b, between BJD~2\,454\,959 and 2\,456\,424. 
It is worth noting that Holczer's catalog lists the TTVs, their associated errors, and the expected mid-transit times from the linear ephemeris. These latter correspond to an average period of $P_b=69.72787813$~days. So in order to work with the observed mid-transit times we must consider the sum between the expected mid-transit times and the TTVs listed in Table 3 of the referred catalog.

\begin{figure*}
    \centering
    \includegraphics[width=\textwidth]{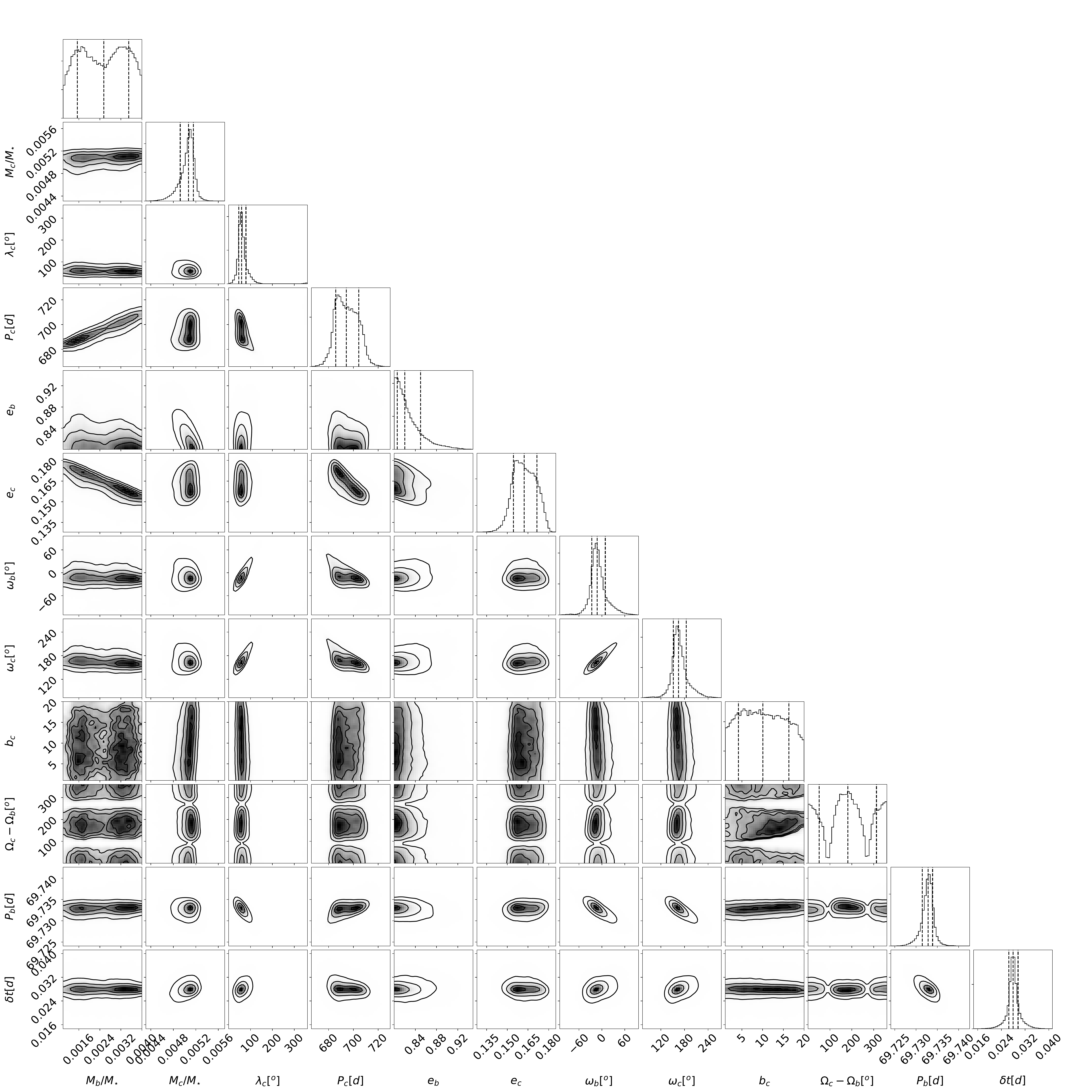}
    \caption{Equally weighted posterior distributions of the 12 parameters of our model (diagonal), and the 
    corresponding correlations between the parameters, for the fit presented in Table~\protect\ref{tab:param419holczerdawson}.}
    \label{fig:triangle419holczer}
\end{figure*}

\begin{figure}
    \centering
	\includegraphics[width=7.5cm,trim={1cm 0 3.5cm 1cm}]{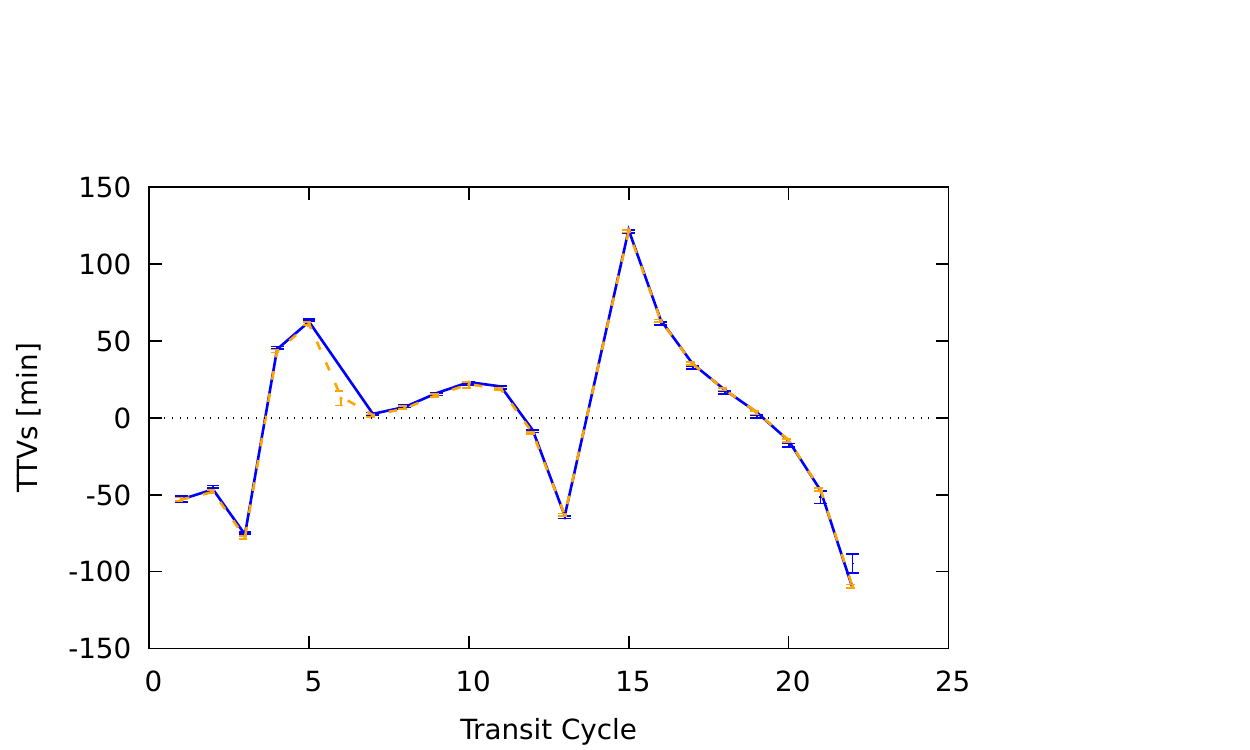}
	\caption {TTVs reported by \protect\citet{Holczer2016, Holczer_Catalog} (blue dots with error bars), and the corresponding best fit presented in Table~\protect\ref{tab:param419holczerdawson} (blue line).
	For comparison, we show the TTVs reported by  
	\protect\citet{Dawson2014} using the \texttt{emcee} package (orange dots with error bars), together with the corresponding best fit obtained in this work (orange dashed line).}
    \label{fig:O-C419holczer}
\end{figure}

\begin{table}
	\caption{Parameters estimated from the best fit to the mid-transit times observed of \protect\citet{Holczer2016}, and their uncertainties, assuming the stellar parameters provided by \protect\citet{Dawson2012}. 
    The orbital parameters are the osculating astrocentric elements at epoch BJD~2\,454\,959.3. The error bars reported for the dynamical fit parameters in are the standard 68.3\% confidence uncertainties.
    The second block reports the derived parameters; $i_p$ is the orbital inclination with respect to the sky plane at transit ($\cos i_p = b_p R_\star /r_p$); $I_p$ is the inclination of the orbit with respect to the transit plane (i.e. the plane where $b=0$); $I_{\mathrm{mut}}$ is the mutual orbital inclination of the orbits.}
	\label{tab:param419holczerdawson}
    \centering  
             \begin{tabular}{lll}
             \hline         
             \hline         
             \multicolumn{3}{| r |}{{\bf{Kepler-419}}  (Dawson et al. 2012)}\\
               \vspace{0.1cm}
               $M_\star$~($M_\odot$) & $1.22 ^{+0.12}_{-0.08}$ & \\
               \vspace{0.1cm}
               $R_\star$~($R_\odot$)  & $1.4^{+0.37}_{-0.21}$ & \\
              \hline
             & \bf{Kepler-419b}  &  \bf{Kepler-419c} \\ [0.5ex] 
              Dynamical fit            &  &     \\ [0.5ex] 
                \vspace{0.1cm}
                $M_p/M_\star\,(\times 10^{-3})$ & $2.54^{+0.95}_{-1.01}$ & $5.06^{+0.08}_{-0.15}$ \\
                \vspace{0.1cm}
                $P_p$~(d) & $69.732^{+0.001}_{-0.001}$ & $694.3^{+10.1}_{-8.6}$ \\
                \vspace{0.1cm}
                $e_p$ & $0.82^{+0.03}_{-0.01}$ & $0.162^{+0.009}_{-0.008}$ \\                
                \vspace{0.1cm}
                $b_p$ & $0.2^{+0.1}_{-0.2}$ & $10.0^{+6.3}_{-5.9}$ \\
                \vspace{0.1cm}                   
                $\varpi_p$~($^\circ$) & $348^{+21}_{-14}$ &  $165^{+19}_{-13}$\\
                \vspace{0.1cm}    
                $\lambda_p$~($^\circ$) & -- & $59^{+20}_{-13}$ \\
                \vspace{0.1cm} 
                $\Omega_p$~($^\circ$) & $270$  & $268^{+142}_{-137}$ \\
                \vspace{0.1cm} 
                $\delta t$~(d) & $0.028^{+0.002}_{-0.001} $  & -- \\    
             \hline         
             Derived parameters    &  &     \\ [0.5ex] 
                \vspace{0.1cm}                
                $M_p$~($M_{\mathrm{Jup}}$) & $ 3.3^{+1.3}_{-1.2} $ & $6.5^{+0.6}_{-0.5}$  \\
                \vspace{0.1cm}                               
                $a_p$~(au) & $0.3545^{+0.0014}_{-0.0009}$ & $1.64^{+0.14}_{-0.09}$ \\     
                \vspace{0.1cm} 
                $i_p$~($^\circ$) & $89.7^{+0.6}_{-0.8}$ & $87.71^{+1.3}_{-1.2}$ \\   
                \vspace{0.1cm}                                 
                $I_p$~($^\circ$) & $1.5^{+0.7}_{-0.9}$ & $1.9^{+1.3}_{-1.2}$ \\   
                \vspace{0.1cm}                                 
                $I_{\mathrm{mut}}$~($^\circ$) & -- & $0.5^{+1.5}_{-0.5}$ \\   
                \vspace{0.1cm}                                 
                $R_p$~($R_{\mathrm{Jup}}$) & $0.8^{+0.2}_{-0.1}$ & -- \\
                \vspace{0.1cm}
                $\rho_p$~(g\,cm$^{-3}$) &  $6.8^{+6.0}_{-3.8}$ & -- \\
             \hline                
               \end{tabular}                
\end{table}

Assuming a star with $1.4\,R_\odot$ and $1.22\,M_\odot$ from \citet{Dawson2012}, and applying \texttt{MultiNest} to the observed mid-transit times, we obtain three possible fits for the model parameters. However, only one of them has a positive value of the logarithm of the Bayesian evidence, $\ln(Z)=75.7$, implying that \texttt{MultiNest} converged to a unique solution. Table~\ref{tab:param419holczerdawson} reports the values of this solution, estimated from the weighted posteriors provided by \texttt{MultiNest}. The distributions of the solution posteriors, and their correlations, are shown in Fig. \ref{fig:triangle419holczer}. The TTVs from Holczer's catalog (blue dots), and the corresponding best fit (solid blue line), are shown in Fig.~\ref{fig:O-C419holczer}. 

The dynamical fit allow us to derived other important physical and orbital planetary parameters (see third part of Table~\ref{tab:param419holczerdawson}). The uncertainties given there are realistic uncertainties obtained by combining those of the stellar parameteres and dynamical fit. For example if we do not consider the uncertainties in the stellar parameters the semi-major axis error bars are of order of $\times10^{-4}$ and $\times10^{-5}$ for each planet.

The non-transiting planet is found to have a mass of $M_c\simeq6.5\, M_\mathrm{Jup}$, a semi-major axis $a_c\simeq1.64$~au, and a moderate eccentricity $e_c\simeq0.16$. The transiting planet has 
$M_b\simeq3.3\, M_\mathrm{Jup}$, $a_b\simeq0.35$~au, and a very high orbital eccentricity, $e_b\simeq0.82$, as expected. 
The two planets are close to the 10:1 mean motion resonance, with a period ratio $P_c/P_b\simeq9.95$, and according to the period uncertainties, they might actually be trapped in this resonance. The orbits are very nearly coplanar, with a mutual inclination of $\simeq0.5^\circ$.

Next, assuming a star with $1.8\,R_\odot$ and $1.39\,M_\odot$ from \citet{Almenara2018}, and applying again \texttt{MultiNest} to the observed mid-transit times calculated from \citet{Holczer_Catalog}, we obtain two possible fits for the model parameters, but only one has a positive value of the logarithm of the evidence, $\ln(Z)=75.6$. Table~\ref{tab:param419holczeralmenara} reports the values of this solution, estimated from the weighted posteriors provided by \texttt{MultiNest}.

\begin{table}
	\caption{Parameters estimated from the best fit to the mid-transit times observed of \protect\citet{Holczer2016}, and their uncertainties, assuming 
	the stellar parameters provided by \protect\cite{Almenara2018}. See Table~\protect\ref{tab:param419holczerdawson} for explanation.}
	\label{tab:param419holczeralmenara}
    \centering  
             \begin{tabular}{lll}
             \hline         
             \hline         
              \multicolumn{3}{| r |}{{\bf{Kepler-419}}  (Almenara et al. 2018)}\\
               \vspace{0.1cm}
               $M_\star$~($M_\odot$) & $1.39^{+0.48}_{-0.48}$ & \\
               \vspace{0.1cm}
               $R_\star$~($R_\odot$)  & $1.8^{+0.2}_{-0.2}$ & \\
              \hline

             & \bf{Kepler-419b}  &  \bf{Kepler-419c} \\ [0.5ex] 
              Dynamical fit            &  &     \\ [0.5ex] 
                \vspace{0.1cm}
                $M_p/M_\star\,(\times 10^{-3})$ & $2.60^{+0.89}_{-1.00}$ & $5.07^{+0.08}_{-0.14}$ \\
                \vspace{0.1cm}
                $P_p$~(d) & $69.7332^{+0.001}_{-0.001}$ & $694.7^{+10.1}_{-8.6}$ \\
                \vspace{0.1cm}
                $e_p$ & $0.82^{+0.02}_{-0.01}$ & $0.162^{+0.009}_{-0.007}$ \\                
                \vspace{0.1cm}
                $b_p$ & $0.26^{+0.11}_{-0.15}$ & $8.9^{+6.3}_{-5.1}$ \\
                \vspace{0.1cm}                   
                $\varpi_p$~($^\circ$) & $346^{+24}_{-12}$ &  $163^{+22}_{-12}$\\
                \vspace{0.1cm}    
                $\lambda_p$~($^\circ$) & -- & $58^{+22}_{-11}$ \\
                \vspace{0.1cm} 
                $\Omega_p$~($^\circ$) & $270$  & $264^{+143}_{-135}$ \\
                \vspace{0.1cm}
                $\delta t$~(d) & $0.028^{+0.001}_{-0.001} $  & -- \\    
             \hline         
             Derived parameters    &  &     \\ [0.5ex] 
                \vspace{0.1cm}                
                $M_p$~($M_{\mathrm{Jup}}$) & $3.8^{+1.8}_{-1.9} $ & $7.4^{+2.5}_{-2.6} $  \\
                \vspace{0.1cm}                               
                $a_p$~(au) & $0.370^{+0.005}_{-0.005} $ & $1.71^{+0.58}_{-0.58} $ \\     
                \vspace{0.1cm} 
                $i_p$~($^\circ$) & $88.1^{+0.8}_{-1.0}$ & $87.8^{+2.1}_{-2.0}$ \\   
                \vspace{0.1cm}                                 
                $I_p$~($^\circ$) & $1.8^{+0.8}_{-1.1}$ & $2.1^{+1.7}_{-1.4}$ \\   
                \vspace{0.1cm}                                 
                $I_{\mathrm{mut}}$~($^\circ$) & -- & $0.4^{+3.3}_{-0.4}$ \\   
                \vspace{0.1cm}                                 
                $R_p$~($R_{\mathrm{Jup}}$) & $1.1^{+0.1}_{-0.1}$ & -- \\
                \vspace{0.1cm}
                $\rho_p$~(g\,cm$^{-3}$) &  $3.7^{+2.3}_{-2.7}$ & -- \\
             \hline                
               \end{tabular}                
\end{table}

This solution provides similar values of the fit parameters to the previous ones. Small differences arise in the masses and semi-major axes: $M_c\simeq7.4\,M_\mathrm{Jup}$, $a_c\simeq1.71$~au, and $M_b\simeq3.8\,M_\mathrm{Jup}$, $a_b\simeq0.37$~au.

The nodal longitue determination of $\Omega_c$ mainly influences the constraint of the planetary orbital mutual inclination ($I_{\mathrm{mut}}$). As we can see in Tables~\ref{tab:param419holczerdawson} and \ref{tab:param419holczeralmenara}, we characterize the nodal longitude with larger uncertainties and this reflects on the bigger mutual inclination uncertainties. We find that $I_{\mathrm{mut}}$ is smaller than $3.7^\circ$ ($1-\sigma$ uncertainty).  

Comparing the 1-$\sigma$ error bars of the parameters presented in the both cases analyzed here we can see how the uncertainties of stellar parameters affect the uncertainties of the derived parameters. Almenara's larger stellar parameters uncertainties results in larger uncertainties of the planetary masses. Beyond this we conclude that both solutions are indistinguishable to within the error bars. Moreover, both solutions have similar values of the Bayesian evidence, implying that \texttt{MultiNest} does not allow us to prefer any solution over the other. 

We then conclude that, independently of the stellar parameters, we can robustly estimate the mass ratio $M_p/M_\star$. On the other hand, the stellar mass value is relevant in the planetary mass determination and might be also relevant to assess the long term stability of the system. We will discuss this possibility in the following section.\\

In order to verify to what extent the mid-transit times data set may introduce any significant bias in the dynamical fit we consider 
\citet{Dawson2014} data. The authors analyzed the light curve of the Kepler-419 system and determined the mid-transit times of Kepler-419b using two different tools: the \texttt{TAP} software \citep[Transit Analysis Package;][]{Gazak2012}, and the \texttt{emcee} package \citep{Foreman2013}. 
Here we apply to the \texttt {emcee} data the same methodology described in Sect.~\ref{methods}. 
We find best fit parameters that are indistinguishable of those presented in Tables~\ref{tab:param419holczerdawson} and \ref{tab:param419holczeralmenara} within their 1-$\sigma$ errors, but interestingly, they show larger values of the evidence: $\ln(Z)=92.17$ assuming Dawson's stellar parameters, and $\ln(Z)=92.01$ assuming Almenara's stellar parameters. This may be due to a more careful treatment of the light curve fit by \citet{Dawson2014} compared to that of \citet{Holczer2016}. However, we find that the TTVs from both Holczer and Dawson agree to within their 1-$\sigma$ confidence levels (blue and orange dots in Fig.~\ref{fig:O-C419holczer}), and the only difference between the two data sets is that \citet{Dawson2014} reported the transit time of one extra cycle. 

\section{Dynamics of the Kepler-419 system}
\label{dynamic}
\begin{figure*}
  \centering
  \includegraphics[width=0.49\textwidth]{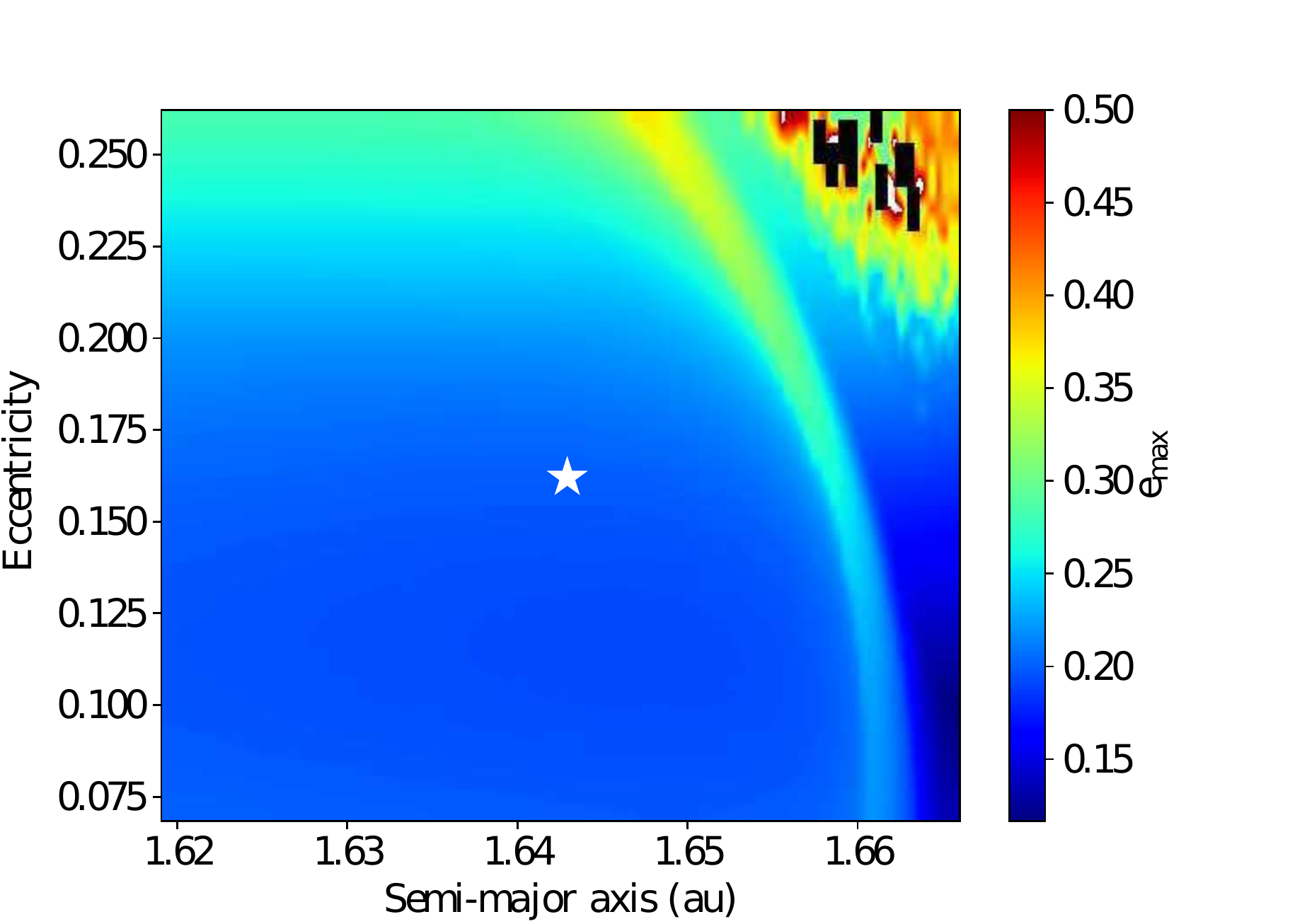}
  \includegraphics[width=0.49\textwidth]{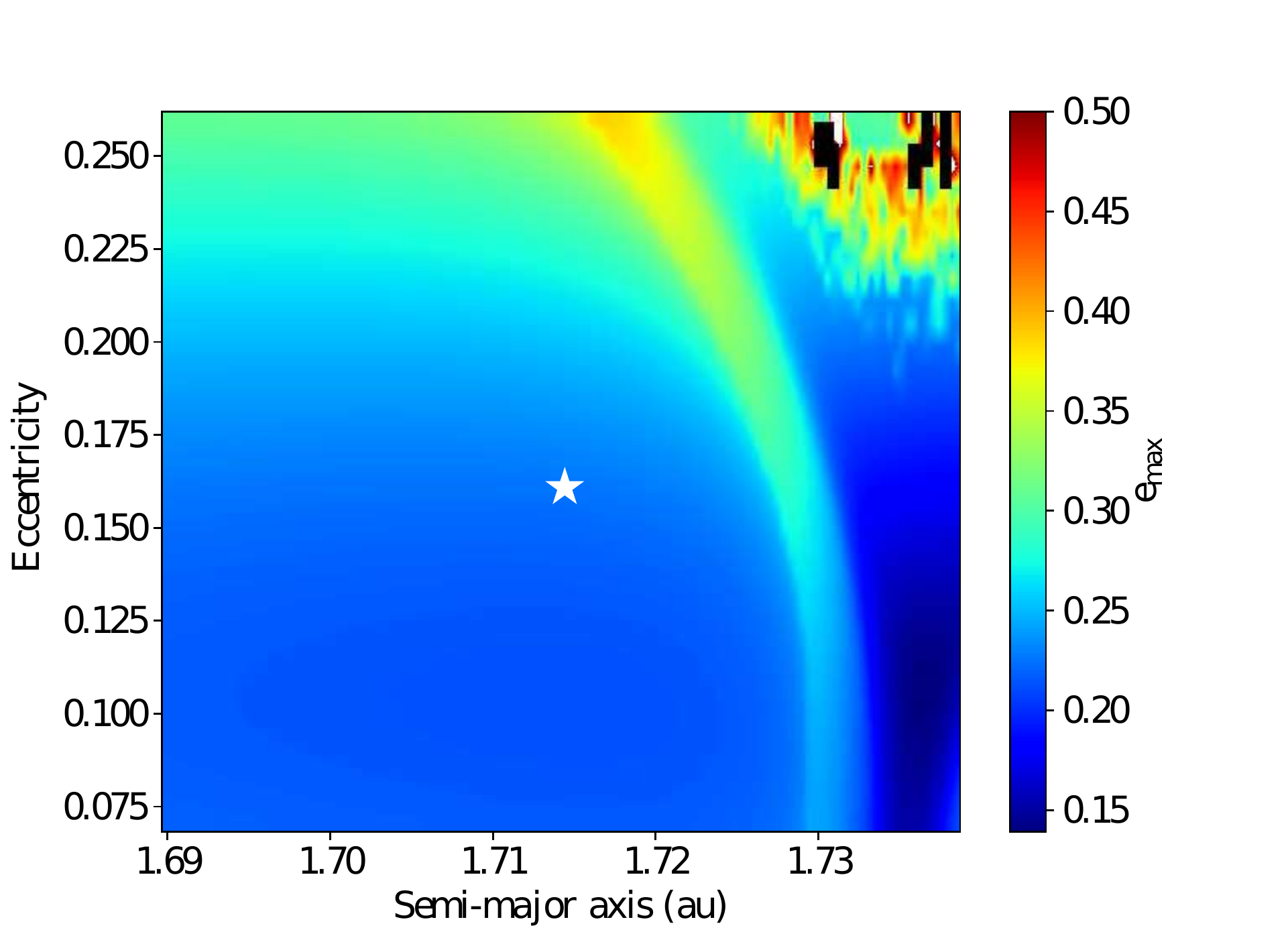} \\
  \includegraphics[width=0.49\textwidth]{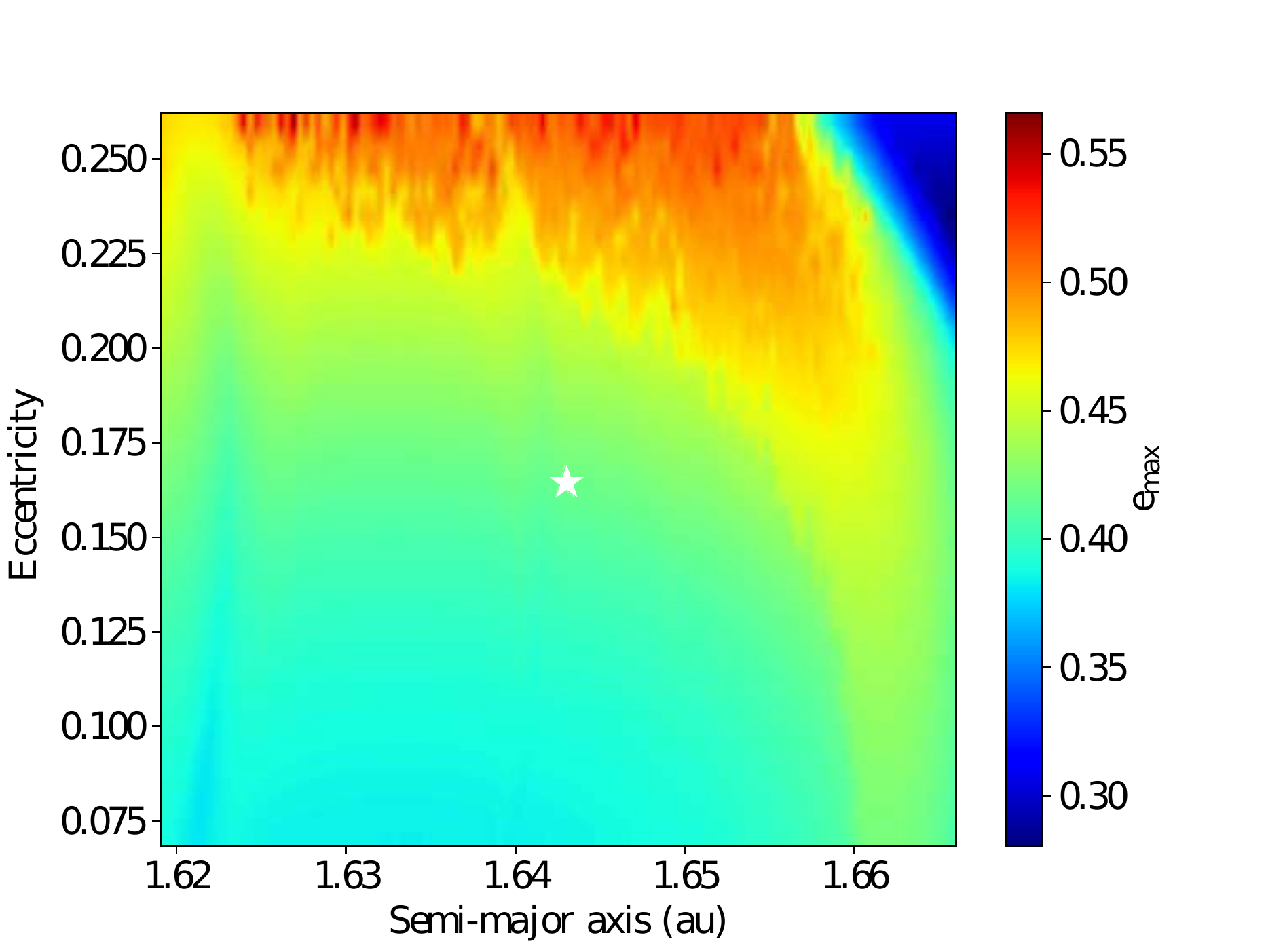}
  \includegraphics[width=0.49\textwidth]{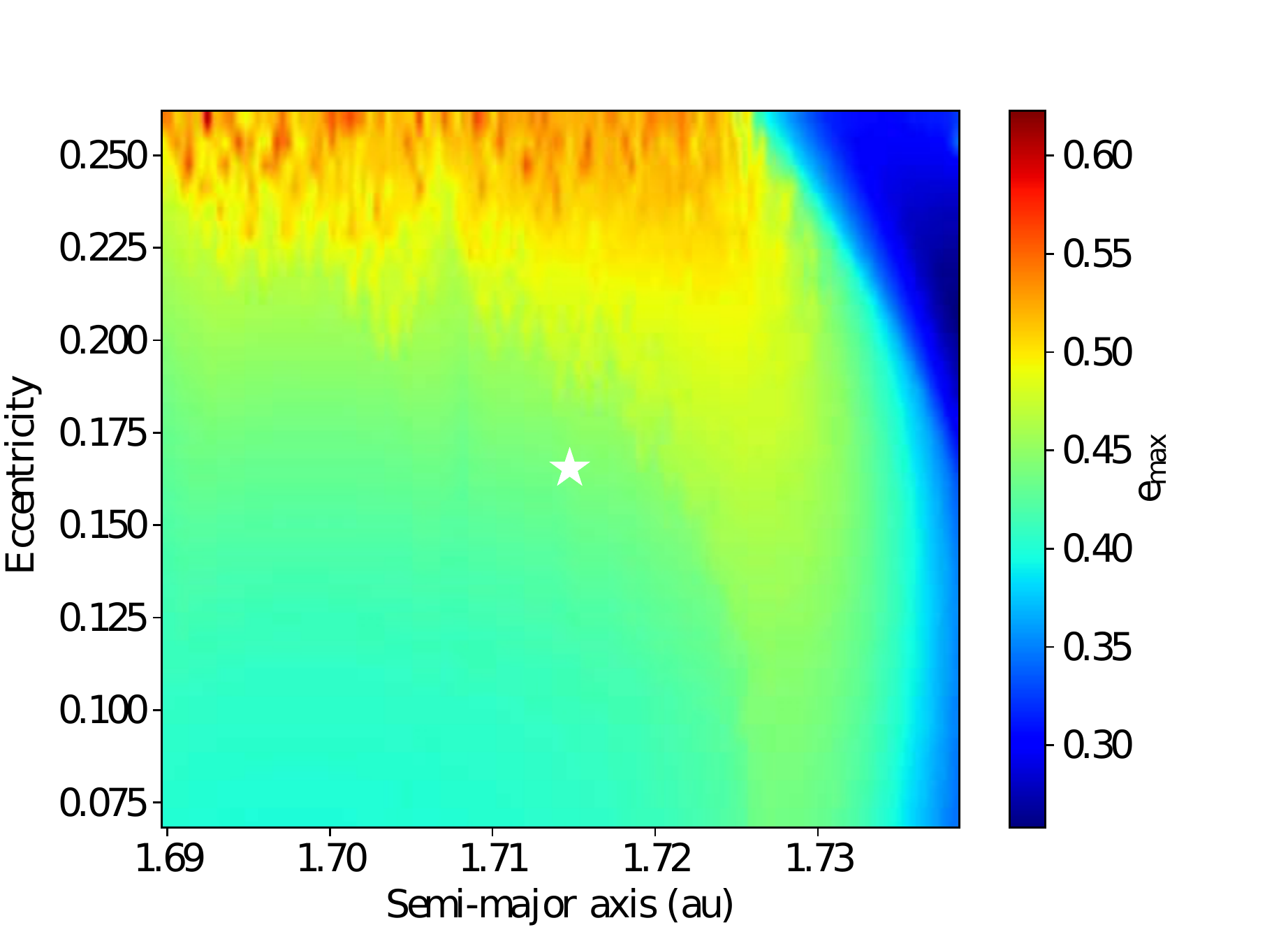} \\
  \includegraphics[width=0.49\textwidth]{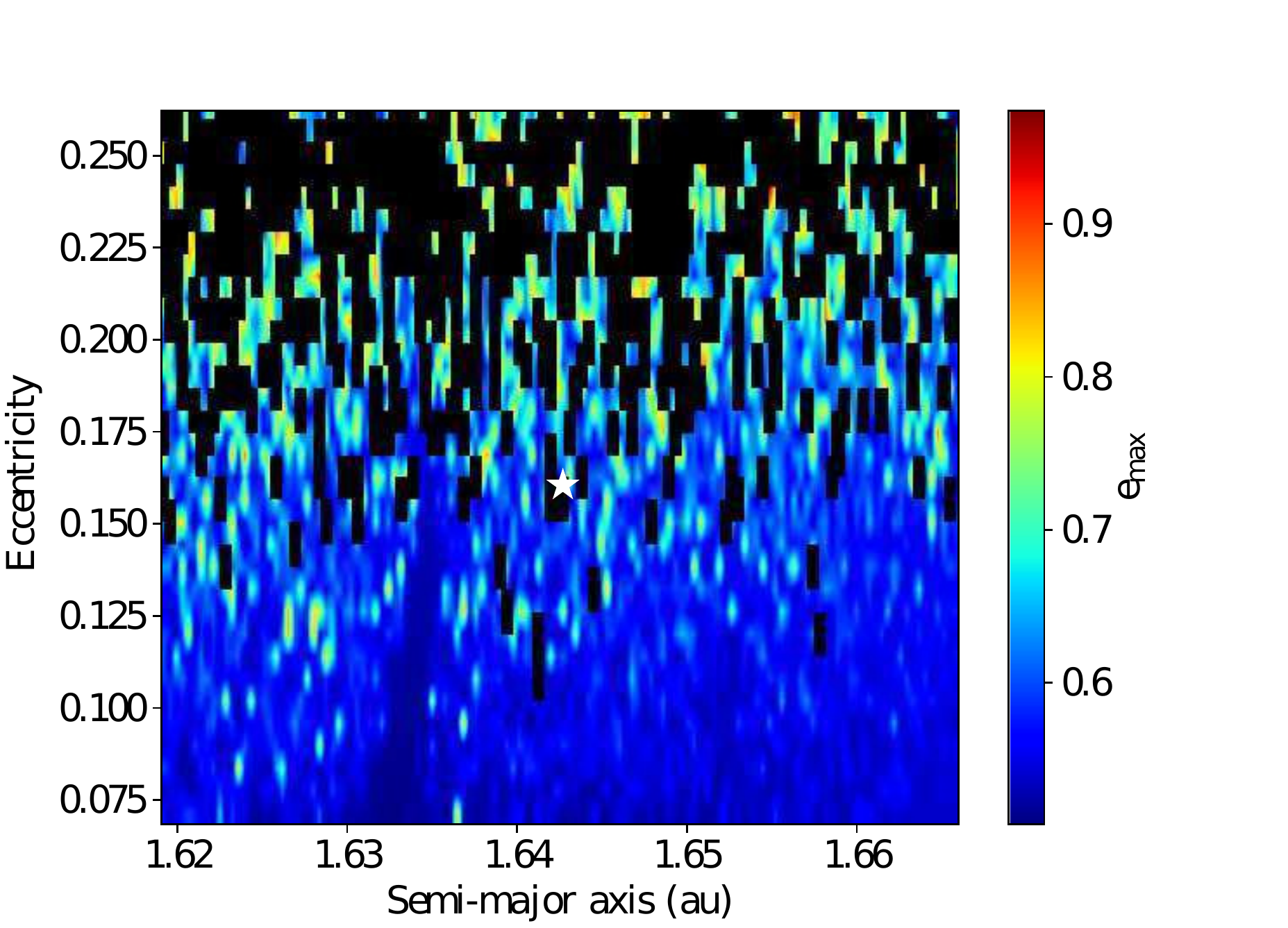}
  \includegraphics[width=0.49\textwidth]{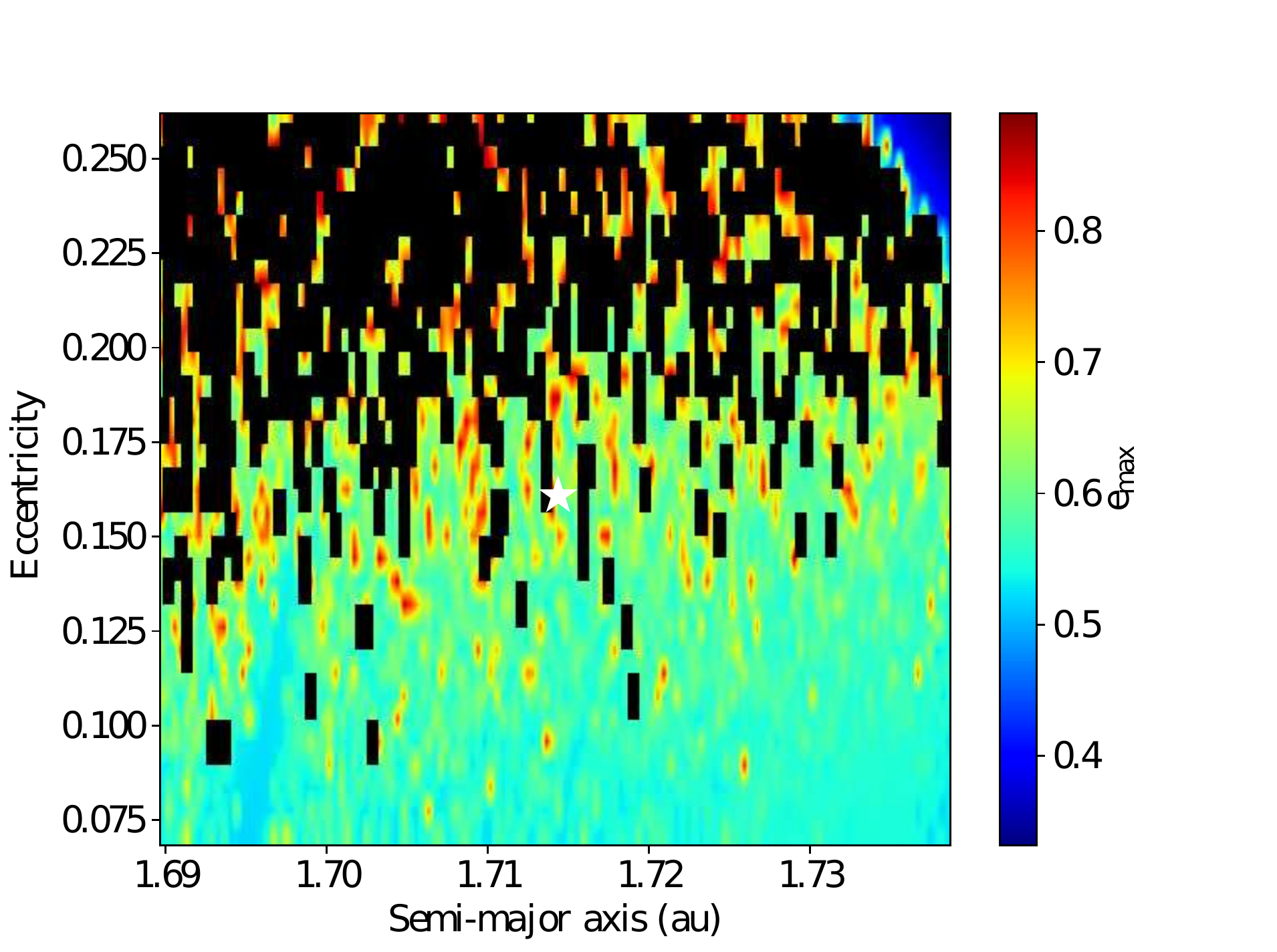}
    \caption{Dynamical maps of the Maximum Eccentricity Index, computed for Kepler-419c. Left column panels correspond to the parameters of Table~\ref{tab:param419holczerdawson}, right column panels correspond to the parameters of Table~\ref{tab:param419holczeralmenara}. Each row corresponds to different values of the mass $M_b$, from the smallest (top) to the largest (bottom) value within the 1-$\sigma$ uncertainties. The best fit parameters of Kepler-419c are indicated by the black dot at the center of the grid. Blue/cyan regions indicate stable motion. Black cells denote unstable initial conditions that either led to a close encounter between the bodies (planet-planet or planet-star), or to the ejection of a planet (see text).}
  \label{fig:maps1Myr}
\end{figure*}

\begin{figure*}
  \centering
  \includegraphics[width=0.47\textwidth]{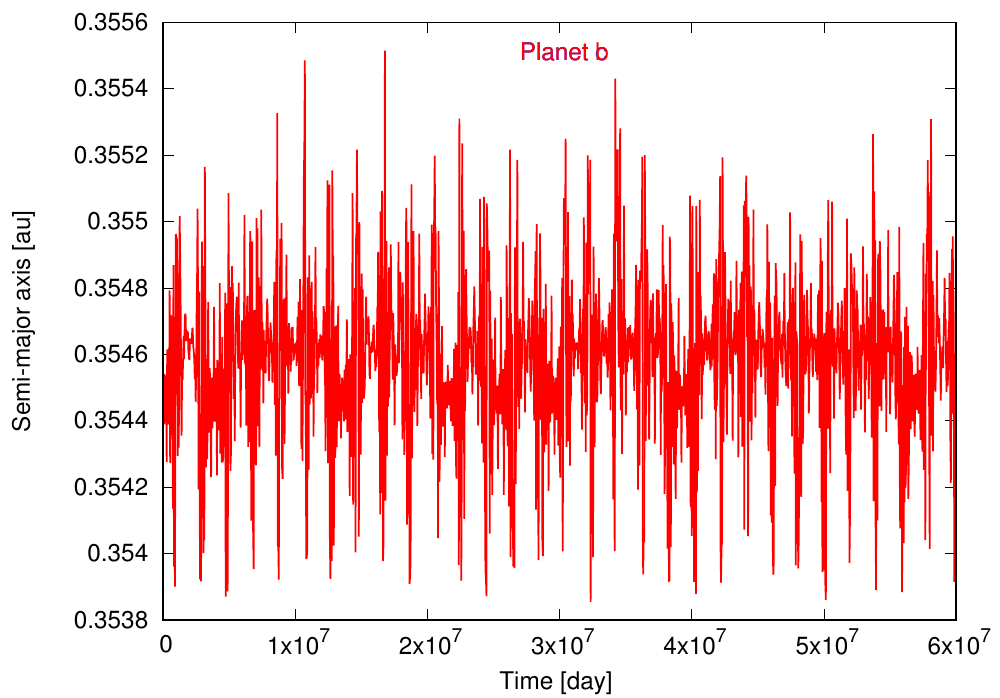}
  \includegraphics[width=0.47\textwidth]{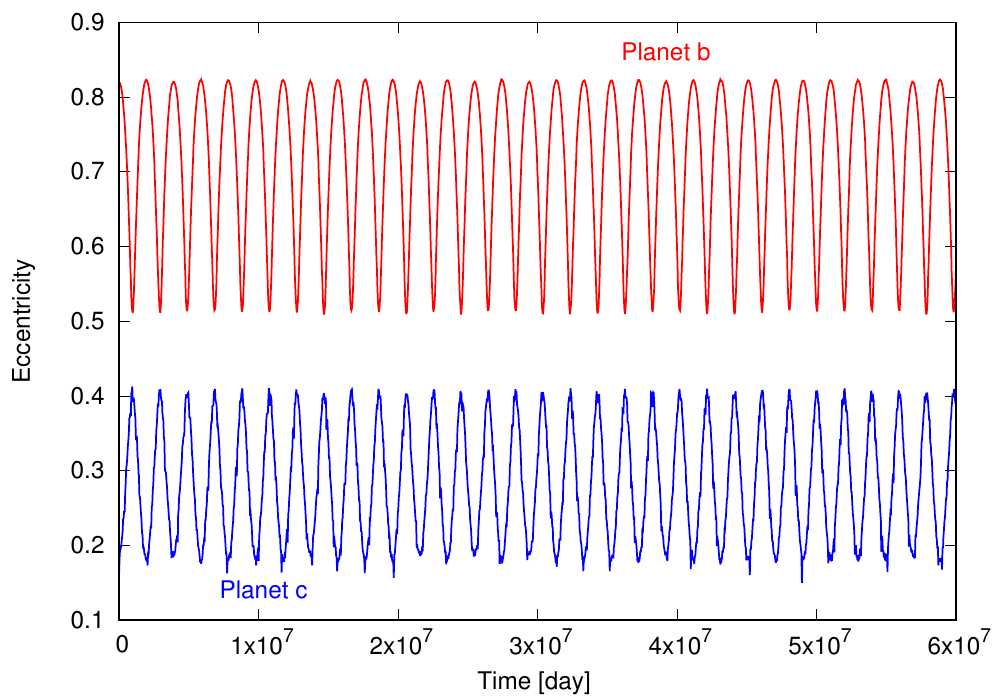}\\
  \includegraphics[width=0.47\textwidth]{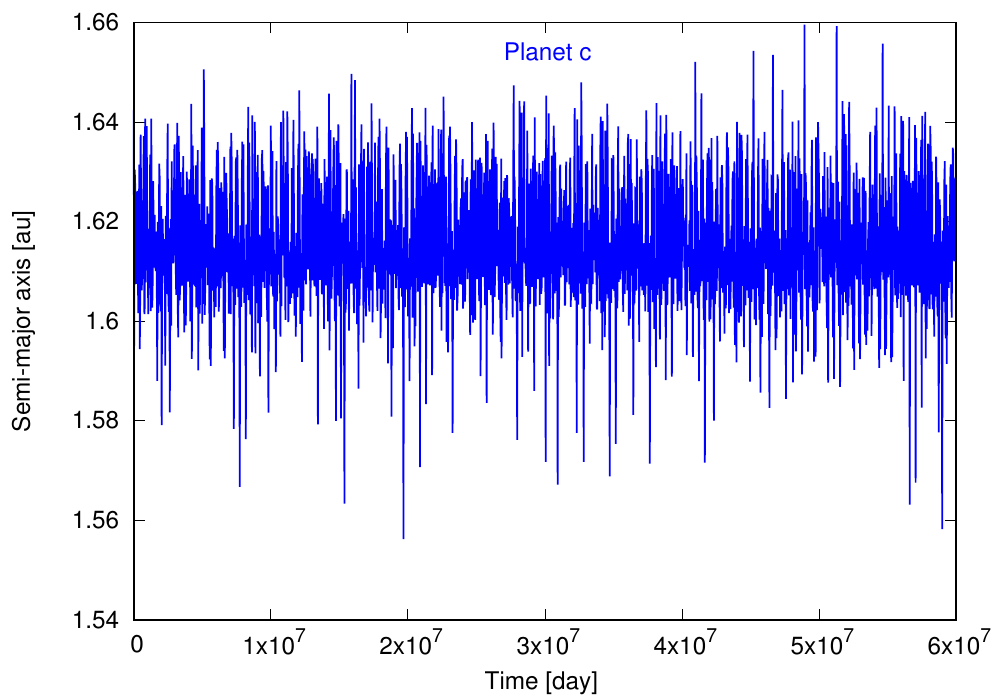} 
  \includegraphics[width=0.47\textwidth]{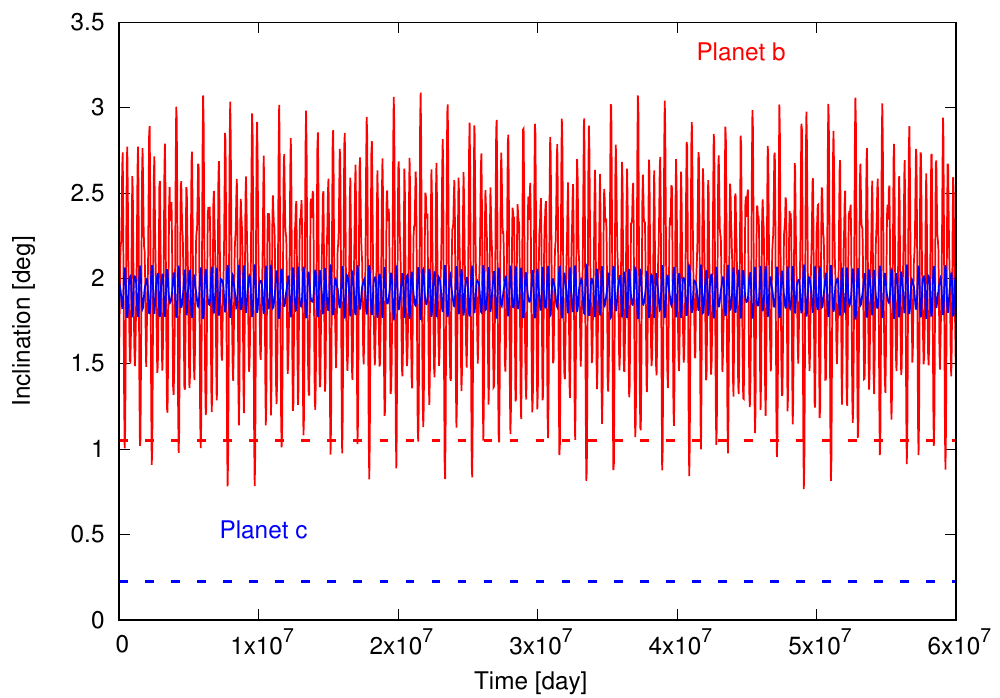} \\
    \caption{Dynamical evolution of the Kepler-419 planets, taking the parameters given in Table~\ref{tab:param419holczerdawson} as initial conditions. Left panels show the evolution of the semi-major axis of planet b (top) and planet c (bottom). The right panels show the evolution of eccentricity (top) and inclination (bottom) of the orbits of planet b (in red) and planet c (in blue). The dotted horizontal lines in the inclination panel correspond to the critical inclination below which it is possible to observe transits.}
  \label{fig:evo1Myr}
\end{figure*}

The very high eccentricity of Kepler-419b, the wide range of possible mass values estimated for this planet, and the proximity of the planets pair to a mean motion resonance, raises the question about the long term stability of this planetary system. \cite{Almenara2018} showed that, for the nominal orbits estimated with their photo-dynamical model, the Kepler-419 system is close to a high eccentricity stable equilibrium point in the ($e_b,\,\varpi_b-\varpi_c$) plane. Here, we explore the dynamics of the system using dynamical maps that span the main uncertainty intervals in the parameters $M_b$ and $a_c$. We choose these parameters because they are not as tightly constrained as the other parameters. 

The maps have been constructed by computing several stability/chaos indicators over a grid of $128\times 32$ initial conditions. The grids span the interval $\pm 15$~days and $\pm0.1$ around the best fit values of $P_c$ and $e_c$. The other orbital elements of the planets are fixed to their best fit values. For each of the solutions presented in Tables~\ref{tab:param419holczerdawson} and \ref{tab:param419holczeralmenara}, we compute three different grids, assuming three different values of the mass $M_b$: one corresponding to the best fit value, and two corresponding to the $\pm 1$-$\sigma$ error. It is worth noting that, since $M_b$ and $P_c$ are strongly correlated (see Fig.~\ref{fig:triangle419holczer}), some combinations of $M_b$ and $a_c$ are outside the $1-\sigma$ region of their joint distribution, and therefore they are missing in our dynamical analisys. The mass of $M_c$ is always fixed to its best fit value. Each initial condition has been simulated over a time span of $5\times 10^{6}$~days, with a time step of 0.2 days, using a version of the \texttt{Swifter\_helio} module, that is part of the \texttt{Swifter} N-body software package (Kaufmann and Levison, \url{www.boulder.swri.edu/swifter/}). This module has been translated into CUDA-C to run on a GPU architecture, which allows us to obtain the maps in only a couple of hours of the computational time (Costa de Souza et al., in preparation).

Figure~\ref{fig:maps1Myr} shows the grids corresponding to the maximum eccentricity index (MEI), which is simply the maximum value of the eccentricity that Kepler-419c attains during the simulation time span. 
This index is straightforward to compute and provides valuable information about the structure of the phase space, easily identifying regions of stability and chaos.

Comparing the panels on the left of Fig.~\ref{fig:maps1Myr} to those on the right, we conclude that the global dynamical behavior of the system is independent of the assumed stellar parameters. On the other hand, the behavior is strongly dependent on the mass of Kepler-419b. Indeed, the larger the mass $M_b$ (top-to-bottom panels), the more unstable the system becomes. Black cells in Fig.~\ref{fig:maps1Myr} indicate initial conditions that led to encounters between the planets within a mutual distance $r<R_b^\mathrm{Hill}+R_c^\mathrm{Hill}$, or to an astrocentric distance $r_p<2R_\star$ or $r_p>10$~au, or to a pericenter distance $q_p<2R_\star$. The corresponding orbits are unstable.

In Fig.~\ref{fig:evo1Myr}, we show the orbital evolution of the planets over $6\times10^7$ days, corresponding to the best fit values from Table~\ref{tab:param419holczerdawson}. We have verified that the behavior is the same over a longer time span of 1~Myr. This evolution has been simulated using the \texttt{Symba} N-body integrator, and considering a 0.2~day time step. We can see that the eccentricities oscillate in opposed phases with very large amplitudes. The same behaviour is observed in the inclinations, although these display smaller amplitudes. In both cases, the opposed phases are related to the conservation of angular momentum of the system. The horizontal dotted lines in the inclinations panel correspond to the inclination value below which the planet will transit the star. Kepler-419b will always display transits, while Kepler-419c will not (at least for the nominal solution).

\section{Discussion and conclusions}
\label{discu}

We compare our best fit parameters of the system to those reported by \citet{Almenara2018}, to find that they are indistinguishable within their 1-$\sigma$ uncertainties. The exception are the angles $\lambda_p$, $\varpi_p$ and $\Omega_p$, that are not directly comparable due to the use of different reference systems. The major difference between Almenara's fit and our fit (Table~\ref{tab:param419holczeralmenara}) is in the estimated mass of Kepler-419b. The value of \citet{Almenara2018} is $M_b/M_{\star}=(1.86\pm 0.25) \times 10^{-3}$, which is close to the lower boundary of the value determined in this work: $M_b/M_{\star}=(2.60^{+0.89}_{-1.00})\times 10^{-3}$. The value of Almenara is also much better constrained, which is expected from the combined use of TTVs and RVs. Almenara's solution would produce a dynamical map that resembles the upper right panel of Fig.~\ref{fig:maps1Myr}. Also, the best fit solution of \citet{Almenara2018} displays a similar behavior as the one shown in Fig.~\ref{fig:evo1Myr}, but the planet eccentricities have more bounded amplitudes. This is related to the smaller mass ratio of Kepler-419b estimated by those authors.

Our main conclusion is that, although the use of TTVs together with RVs certainly helps to improve and constrain the parameters of a multiplanetary system, the use alone of a large number of TTVs, with a high sinal to noise ratio, is still capable to provide quite confident and robust estimates 
of the parameters, as we can see in this work. This implies that we can exploit the availability of the large database of TTVs of \citet{Holczer2016} to characterize many planetary systems around faint Kepler stars, for which the RVs measurements are not available.

\section*{Acknowledgements}

The simulations have been performed at the SDumont cluster
of the Brazilian System of High Performance Computing (SINAPAD). This work has been supported
by the Brazilian National Council of Research (CNPq), by the Brazilian Federal Agency for Support and Assessment of Postgraduate Education from the Ministry of Education (CAPES) and by NASA's XRP program.
 


\bibliographystyle{mnras}
\bibliography{kep419} 

\begin{thebibliography}{}
\makeatletter
\relax
\def\mn@urlcharsother{\let\do\@makeother \do\$\do\&\do\#\do\^\do\_\do\%\do\~}
\def\mn@doi{\begingroup\mn@urlcharsother \@ifnextchar [ {\mn@doi@}
  {\mn@doi@[]}}
\def\mn@doi@[#1]#2{\def\@tempa{#1}\ifx\@tempa\@empty \href
  {http://dx.doi.org/#2} {doi:#2}\else \href {http://dx.doi.org/#2} {#1}\fi
  \endgroup}
\def\mn@eprint#1#2{\mn@eprint@#1:#2::\@nil}
\def\mn@eprint@arXiv#1{\href {http://arxiv.org/abs/#1} {{\tt arXiv:#1}}}
\def\mn@eprint@dblp#1{\href {http://dblp.uni-trier.de/rec/bibtex/#1.xml}
  {dblp:#1}}
\def\mn@eprint@#1:#2:#3:#4\@nil{\def\@tempa {#1}\def\@tempb {#2}\def\@tempc
  {#3}\ifx \@tempc \@empty \let \@tempc \@tempb \let \@tempb \@tempa \fi \ifx
  \@tempb \@empty \def\@tempb {arXiv}\fi \@ifundefined
  {mn@eprint@\@tempb}{\@tempb:\@tempc}{\expandafter \expandafter \csname
  mn@eprint@\@tempb\endcsname \expandafter{\@tempc}}}

\bibitem[\protect\citeauthoryear{{Agol}, {Steffen}, {Sari}  \&
  {Clarkson}}{{Agol} et~al.}{2005}]{Agol2005}
{Agol} E.,  {Steffen} J.,  {Sari} R.,   {Clarkson} W.,  2005, \mn@doi [\mnras]
  {10.1111/j.1365-2966.2005.08922.x}, \href
  {http://adsabs.harvard.edu/abs/2005MNRAS.359..567A} {359, 567}

\bibitem[\protect\citeauthoryear{{Almenara} et~al.,}{{Almenara}
  et~al.}{2018}]{Almenara2018}
{Almenara} J.~M.,  et~al., 2018, \mn@doi [A\&A] {10.1051/0004-6361/201732500},
  615, A90

\bibitem[\protect\citeauthoryear{{Ballard} et~al.,}{{Ballard}
  et~al.}{2011}]{Ballard2011}
{Ballard} S.,  et~al., 2011, \mn@doi [\apj] {10.1088/0004-637X/743/2/200},
  \href {http://adsabs.harvard.edu/abs/2011ApJ...743..200B} {743, 200}

\bibitem[\protect\citeauthoryear{{Borucki} et~al.,}{{Borucki}
  et~al.}{2011}]{Borucki2011b}
{Borucki} W.~J.,  et~al., 2011, \mn@doi [\apj] {10.1088/0004-637X/736/1/19},
  \href {http://adsabs.harvard.edu/abs/2011ApJ...736...19B} {736, 19}

\bibitem[\protect\citeauthoryear{Dawson \& Johnson}{Dawson \&
  Johnson}{2012}]{DawsonJohnson}
Dawson R.~I.,  Johnson J.~A.,  2012, \apj, 756, 122

\bibitem[\protect\citeauthoryear{{Dawson}, {Johnson}, {Morton}, {Crepp},
  {Fabrycky}, {Murray-Clay}  \& {Howard}}{{Dawson} et~al.}{2012}]{Dawson2012}
{Dawson} R.~I.,  {Johnson} J.~A.,  {Morton} T.~D.,  {Crepp} J.~R.,  {Fabrycky}
  D.~C.,  {Murray-Clay} R.~A.,   {Howard} A.~W.,  2012, \mn@doi [\apj]
  {10.1088/0004-637X/761/2/163}, \href
  {http://adsabs.harvard.edu/abs/2012ApJ...761..163D} {761, 163}

\bibitem[\protect\citeauthoryear{{Dawson} et~al.,}{{Dawson}
  et~al.}{2014}]{Dawson2014}
{Dawson} R.~I.,  et~al., 2014, \mn@doi [\apj] {10.1088/0004-637X/791/2/89},
  \href {http://adsabs.harvard.edu/abs/2014ApJ...791...89D} {791, 89}

\bibitem[\protect\citeauthoryear{{Deck}, {Agol}, {Holman}  \&
  {Nesvorn{\'y}}}{{Deck} et~al.}{2014}]{Deck2014}
{Deck} K.~M.,  {Agol} E.,  {Holman} M.~J.,   {Nesvorn{\'y}} D.,  2014, \mn@doi
  [\apj] {10.1088/0004-637X/787/2/132}, \href
  {http://adsabs.harvard.edu/abs/2014ApJ...787..132D} {787, 132}

\bibitem[\protect\citeauthoryear{{Feroz}, {Hobson}  \& {Bridges}}{{Feroz}
  et~al.}{2009}]{Feroz2009}
{Feroz} F.,  {Hobson} M.~P.,   {Bridges} M.,  2009, \mn@doi [\mnras]
  {10.1111/j.1365-2966.2009.14548.x}, \href
  {http://adsabs.harvard.edu/abs/2009MNRAS.398.1601F} {398, 1601}

\bibitem[\protect\citeauthoryear{{Feroz}, {Hobson}, {Cameron}  \&
  {Pettitt}}{{Feroz} et~al.}{2013}]{Feroz2013}
{Feroz} F.,  {Hobson} M.~P.,  {Cameron} E.,   {Pettitt} A.~N.,  2013, preprint,
  \href {http://adsabs.harvard.edu/abs/2013arXiv1306.2144F} {} (\mn@eprint
  {arXiv} {1306.2144})

\bibitem[\protect\citeauthoryear{Foreman-Mackey, Hogg, Lang  \&
  Goodman}{Foreman-Mackey et~al.}{2013}]{Foreman2013}
Foreman-Mackey D.,  Hogg D.~W.,  Lang D.,   Goodman J.,  2013, Publications of
  the Astronomical Society of the Pacific, 125, 306

\bibitem[\protect\citeauthoryear{{Gazak}, {Johnson}, {Tonry}, {Dragomir},
  {Eastman}, {Mann}  \& {Agol}}{{Gazak} et~al.}{2012}]{Gazak2012}
{Gazak} J.~Z.,  {Johnson} J.~A.,  {Tonry} J.,  {Dragomir} D.,  {Eastman} J.,
  {Mann} A.~W.,   {Agol} E.,  2012, \mn@doi [Advances in Astronomy]
  {10.1155/2012/697967}, \href
  {http://cdsads.u-strasbg.fr/abs/2012AdAst2012E..30G} {2012, 697967}

\bibitem[\protect\citeauthoryear{{Holczer} et~al.,}{{Holczer}
  et~al.}{2016a}]{Holczer_Catalog}
{Holczer} T.,  et~al., 2016a, VizieR Online Data Catalog, \href
  {http://adsabs.harvard.edu/abs/2016yCat..22250009H} {222}

\bibitem[\protect\citeauthoryear{{Holczer} et~al.,}{{Holczer}
  et~al.}{2016b}]{Holczer2016}
{Holczer} T.,  et~al., 2016b, \mn@doi [\apjs] {10.3847/0067-0049/225/1/9},
  \href {http://adsabs.harvard.edu/abs/2016ApJS..225....9H} {225, 9}

\bibitem[\protect\citeauthoryear{{Holman} \& {Murray}}{{Holman} \&
  {Murray}}{2005}]{Holman2005}
{Holman} M.~J.,  {Murray} N.~W.,  2005, \mn@doi [Science]
  {10.1126/science.1107822}, \href
  {http://adsabs.harvard.edu/abs/2005Sci...307.1288H} {307, 1288}

\bibitem[\protect\citeauthoryear{{Holman} et~al.,}{{Holman}
  et~al.}{2010}]{Holman2010}
{Holman} M.~J.,  et~al., 2010, \mn@doi [Science] {10.1126/science.1195778},
  \href {http://adsabs.harvard.edu/abs/2010Sci...330...51H} {330, 51}

\bibitem[\protect\citeauthoryear{{Levison} \& {Duncan}}{{Levison} \&
  {Duncan}}{1994}]{Levison1994}
{Levison} H.~F.,  {Duncan} M.~J.,  1994, \mn@doi [\icarus]
  {10.1006/icar.1994.1039}, \href
  {http://adsabs.harvard.edu/abs/1994Icar..108...18L} {108, 18}

\bibitem[\protect\citeauthoryear{{Nesvorn{\'y}}, {Kipping}, {Buchhave},
  {Bakos}, {Hartman}  \& {Schmitt}}{{Nesvorn{\'y}} et~al.}{2012}]{Nesvorny2012}
{Nesvorn{\'y}} D.,  {Kipping} D.~M.,  {Buchhave} L.~A.,  {Bakos} G.~{\'A}.,
  {Hartman} J.,   {Schmitt} A.~R.,  2012, \mn@doi [Science]
  {10.1126/science.1221141}, \href
  {http://adsabs.harvard.edu/abs/2012Sci...336.1133N} {336, 1133}

\bibitem[\protect\citeauthoryear{{Nesvorn{\'y}}, {Kipping}, {Terrell},
  {Hartman}, {Bakos}  \& {Buchhave}}{{Nesvorn{\'y}}
  et~al.}{2013}]{Nesvorny2013}
{Nesvorn{\'y}} D.,  {Kipping} D.,  {Terrell} D.,  {Hartman} J.,  {Bakos}
  G.~{\'A}.,   {Buchhave} L.~A.,  2013, \mn@doi [\apj]
  {10.1088/0004-637X/777/1/3}, \href
  {http://adsabs.harvard.edu/abs/2013ApJ...777....3N} {777, 3}

\bibitem[\protect\citeauthoryear{{Saad-Olivera}, {Nesvorn{\'y}}, {Kipping}  \&
  {Roig}}{{Saad-Olivera} et~al.}{2017}]{Saad-Olivera2017}
{Saad-Olivera} X.,  {Nesvorn{\'y}} D.,  {Kipping} D.~M.,   {Roig} F.,  2017,
  \mn@doi [\aj] {10.3847/1538-3881/aa64e0}, \href
  {http://adsabs.harvard.edu/abs/2017AJ....153..198S} {153, 198}

\makeatother
\end{thebibliography}


\bsp	
\label{lastpage}
\end{document}